\documentstyle[aps,prl,psfig,epsfig,twocolumn,psfrag]{revtex}
\input{epsf}
\draft
\begin{document}
\title{Wave scattering by discrete breathers}
\author{S. Flach, A. E.
Miroshnichenko and M. V. Fistul\footnotemark[2]}
\address{
Max-Planck-Institut f\"ur Physik komplexer Systeme, N\"othnitzer
Strasse 38, D-01187 Dresden, Germany}

\date{\today}
\wideabs{
\maketitle
\begin{abstract}
We present a theoretical study of linear wave scattering in one-dimensional
nonlinear lattices by {\it intrinsic spatially localized dynamic} excitations
or {\it discrete breathers}.
These states appear in various nonlinear systems and present
a time-periodic localized scattering potential for plane waves.
We consider the case of elastic one-channel scattering,
when the frequencies of
incoming and transmitted waves coincide, but the breather provides with
additional spatially localized ac channels whose presence may lead to various
interference patterns.
The dependence of the transmission coefficient on the wave number $q$
and the breather frequency
$\Omega_b$ is studied for different types of breathers: acoustic and
optical breathers, and
rotobreathers. We identify
several typical scattering setups where the internal time dependence of the
breather is of crucial importance for the observed transmission properties.
\end{abstract}
}
\footnotetext[2]{Present address: Physikalisches Institut III,
Universit\"at Erlangen-N\"urnberg, D-91058, Erlangen, Germany}

\section{Introduction}

The problem of wave propagation through media with various
inhomogeneities has been a complex issue of constant interest and
appears in different areas of physics. Particular examples are
acoustic and electromagnetic wave propagation in various
disordered media \cite{psbwzzgp89,Klyatskin}, tunneling of
electrons in solids\cite{LifGredPast} and electron transport
through quantum (molecular) wires \cite{Zwerger,sagybl93}. In many
cases of interest the conductivity (electron transport)
\cite{LifGredPast,sagybl93,pjp93} and the heat conductivity
(phonon transport)\cite{savgov01,slrlap97,slrlap98} are determined
through the wave scattering by spatial inhomogeneities. Of
particular interest is the wave propagation in one-dimensional
systems where interference effects may be strongly enhanced.

In most of the studies wave scattering by {\it static localized}
inhomogeneities has been considered. More recently the scattering
by generic time-dependent potentials has received strong attention
\cite{IVMel,Lemp,Hangi}. This is due to the possibility to
generate various time-dependent scattering potentials artificially
e.g. in the presence of laser beams or microwave radiation.
Several interesting effects such as giant enhancement of
tunneling\cite{IVMel,Lemp} and Fano
resonances\cite{uf61,junads94,cskams98,swksk01} have been found.

It is a well established fact that various {\it nonlinear
spatially discrete} systems can support different types of
excitations, namely, propagating linear waves ({\it phonons}) and
time-periodic spatially localized excitations called discrete {\it
breather} states (DB) \cite{ajsjbp95,sa97,sfcrw98}. The origin of
the latter localized states is not the presence of disorder but
rather the peculiar interplay between the nonlinearity and
discreteness of the lattice. While the nonlinearity provides with
an amplitude-dependent tunability of oscillation or rotation
frequencies of DBs, $\Omega_b$, the spatial discreteness of the
system leads to finite upper bounds of the frequency spectrum of
small amplitude waves $\omega_q$. It allows to escape resonances
of all multiples of the breather frequency $\Omega_b$ with
$\omega_q$. Note here, that nonlinear discrete lattices admit
different types of DBs depending on the spectrum of linear waves
propagating in the lattice, i. e. acoustic breathers and
rotobreathers (Fig. 1a and 1b), optical breathers (Fig. 1c), etc.
Such properties of DBs as their frequency dependent localization
length and the stability of DBs with respect to small amplitude
perturbations have been widely studied. DBs have been observed in
experiments covering such diverse fields as nonlinear optics
\cite{hseysrmarbjsa98}, interacting Josephson junctions
\cite{etjjmtpo00,pbdaavusfyz00} magnetic systems\cite{utslqeajs99}
and lattice dynamics of crystals \cite{arbdccpslbhsrp84}.

%
\begin{figure}[htb]
\vspace{20pt}
\centerline{\psfig{figure=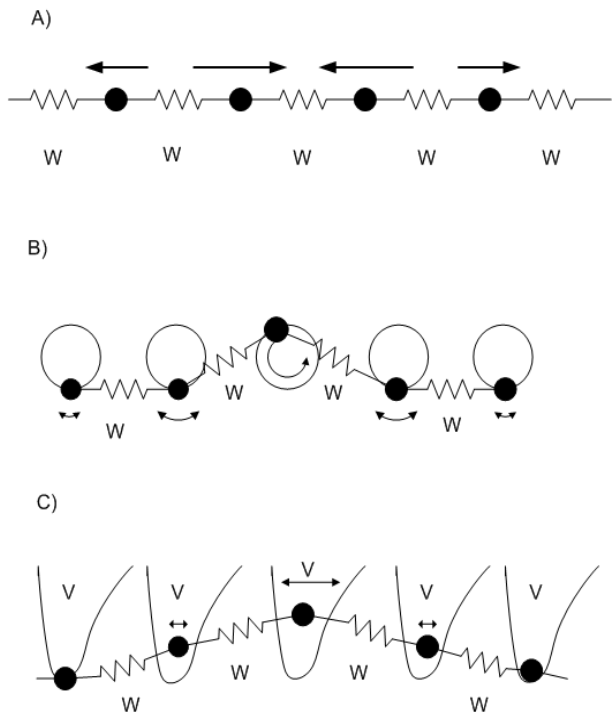,width=82mm,height=95mm}}
\vspace{2pt}
\caption{A schematic representation of different types of discrete breathers:
a) acoustic breather; b) acoustic rotobreather;
c) optical breather.}
\label{fig1}
\end{figure}

Although DBs present complex dynamical objects, in many cases
experimental measurements can be well understood by using certain
{\it time-averaged} properties of DBs
\cite{etjjmabtpo00,aemsfmvfyzjbp01}. Thus, a natural question
appears whether the internal breather dynamics is of crucial
importance to understand the lattice dynamics in the presence of
DBs. In this paper we study the propagation of small amplitude
plane waves in one-dimensional nonlinear lattices in the presence
of a DB and obtain that the internal dynamics of the DB may lead
to a drastical increase or decrease of wave transmission as
compared to the time-averaged setup. Thus the wave scattering by
DBs is interesting both as some spectroscopical tool to study  DB
properties and as a way to control the wave transmission
(conductivity) by varying the DB state. Finally our studies are of
use for the general understanding of wave scattering by
time-periodic potentials.

First successful attempts to describe the variety of phenomena
arising from wave scattering by DBs have been performed some time
ago \cite{sfcw95,tcsasf98,swksk00,skcbrsm97,swksk01}. While a
number of results have been obtained, we are far from a full
description of the complexity of possible scattering outcomes.
This concerns the wave scattering in systems with acoustic spectra
$\omega_q$, the importance of the internal DB dynamics including
the comparison between rotational and vibrational excitations, the
dependence of the wave propagation on the DB energy, and the
single channel elastic versus the two channel inelastic scattering
cases\cite{tcsasf98}. Here we will address these problems but
restrict  ourselves to the elastic scattering case.

The paper is organized as follows:
in Section II we present the general formalism and describe analytical
and novel numerical methods
used to analyze the wave scattering by DBs, in Sections III, IV and V the dependencies of
the transmission coefficient on the breather frequency $\Omega_b$ and the  wave number $q$
for different types of breathers (acoustic breathers and rotobreathers, optical breathers,
see Fig. 1) are obtained, and finally a discussion is provided in Section VI.

\section{General formalism}

To proceed we will consider
one-dimensional nonlinear lattices with nearest neighbor interaction,
optional on-site (substrate) potential, and with one degree of freedom per lattice site.
Both the increasing of the interaction range and the extension
to more than the one degree of freedom per lattice site are not of crucial importance.

The dynamics of the system is characterized by time-dependent coordinates $x_n(t)$ and
the class of Hamiltonians considered here reads
\begin{eqnarray}
{\mathrm H}=\sum_{n}\left( \frac{\dot{x}_n^2}{2}+V[x_n]+W[x_n-x_{n-1}] \right)\;.
\end{eqnarray}
Here $V[x]$ is an optional
on-site (substrate) potential and $W[x]$ is the nearest neighbor
interaction.
The equations of motion become
\begin{eqnarray}\label{Mot}
\ddot{x}_n=-W^{\prime}[x_n-x_{n-1}]+W^{\prime}[x_{n+1}-x_n]-V^{\prime}[x_n]~~.
\end{eqnarray}
Without loss of generality we take $V[0]=W[0]=V'[0]=W'[0]=0$
and $V''[0] \geq 0$, $W''[0] > 0$.
This Hamiltonian supports the excitation of small amplitude linear waves with
the frequency spectrum
\begin{equation}\label{spectr}
\omega_q^2 = V''[0] + 4W''[0]\sin^2\left( \frac{q}{2} \right)~~,
\end{equation}
with $q$ being the wave number.

Time-dependent spatially localized solutions (DBs) exist
for different types of potentials $V[x]$ and $W[x]$, although
at least one of the two functions ($V[x]$ and/or $W[x]$) has to be anharmonic.
DB solutions are
characterized by being time-periodic
$\hat{x}_n(t+T_b)=\hat{x}_n(t)$ and spatially localized
$\hat{x}_{|n| \rightarrow \infty} \rightarrow 0$ (except systems with
$V=0$ where $\hat{x}_{n \rightarrow \pm \infty} \rightarrow \pm c$
with $c$ possibly being nonzero). 
If the Hamiltonian $H$ is invariant under a finite
translation (rotation) of any $x_n \rightarrow x_n + \lambda$, then
discrete {\it rotobreathers} (DRB) may exist \cite{stmp96}. These excitations are characterized by
one or several
sites in the breather center evolving in a rotational state
$\hat{x}_0(t+T_b)=\hat{x}_0(t)+\lambda$, while outside this center
the lattice is governed again by
time periodic spatially localized oscillations.
The breather frequency $\Omega_b = 2\pi/T_b$ can generally take any
values provided $k\Omega_b \neq \omega_q$ for all nonzero integer $k$.
As $\omega_q^2$ is an analytic function of $q$, DBs are exponentially
localized on the lattice.

\subsection{The linearized problem}

To study the
scattering of small amplitude plane waves by a DB we
linearize the equations of
motion (\ref{Mot}) around a breather solution $x_n(t)=\hat{x}_n(t) + \epsilon_n(t)$:
\begin{eqnarray}\label{scat1}
\ddot{\epsilon}_n&=&-W^{\prime\prime}[\hat{x}_n(t)
-\hat{x}_{n-1}(t)](\epsilon_n-\epsilon_{n-1})\nonumber\\
&&+W^{\prime\prime}[\hat{x}_{n+1}(t)-\hat{x}_n(t)](\epsilon_{n+1}
-\epsilon_n)-V^{\prime\prime}[\hat{x}_n(t)]\epsilon_n\;.
\end{eqnarray}
This is a set of coupled linear differential equations
with time periodic coefficients of period $T_b$. Note that these coefficients
are determined by the given DB solution $\hat{x}_n(t)$.

Eq.(\ref{scat1}) determines the linear stability of the breather through the
spectral
properties of the Floquet matrix,\cite{sfcrw98,tcsasf98} which is given by a map over
one breather period
\begin{eqnarray}\label{Floquet}
\left(\begin{array}{l}\vec{\epsilon}(T_b)\\\dot{\vec{\epsilon}}(T_b) \end{array}
\right)={\mathbf F}\left(\begin{array}{l}\vec{\epsilon}(0)\\\dot{\vec{\epsilon}}(0)
\end{array} \right),
\end{eqnarray}
where
$\vec{\epsilon} \equiv (... ,\epsilon_{n-1},\epsilon_n,\epsilon_{n+1},...)$.
For marginally stable breathers
all eigenvalues of the symplectic matrix ${\mathbf F}$ will be
located on the unit
circle and can be written as $e^{i\theta}$.
The corresponding eigenvectors are the
solutions of Eq.(\ref{scat1}), and fulfill the Bloch condition
\begin{eqnarray}\label{Bloch}
\epsilon_{n}(t+T_b)=e^{-i\theta t/T_b}\epsilon_{n}(t)\;.
\end{eqnarray}
Because the DB solution is exponentially localized on the lattice,
equation (\ref{scat1}) will reduce to the usual small amplitude wave
equation far away from the breather center. Thus, only a finite number
of Floquet eigenvectors are spatially localized, while an infinite
number of them (for an infinite lattice) are delocalized and these solutions correspond
to plane waves with the spectrum (\ref{spectr}) and the Floquet
phases $\theta=\pm\omega_qT_b$.
The remaining Floquet eigenvalues
correspond to local Floquet eigenvectors and are separated from the
plane wave spectrum on the unit circle.
Note here, that two eigenvectors with the degenerated eigenvalue
$e^{i\theta}~=~+1$ always exist and reflect perturbations tangent to the breather
family of solutions.

As a consequence of the Bloch condition (\ref{Bloch}) any spatially extended
Floquet eigenvector can be represented in the form
\begin{equation}
\epsilon_n(t) = \sum_{k=-\infty}^{\infty}
e_{nk} {\rm e}^{{\rm i}(\omega_q + k\Omega_b)t}.
\label{Bloch2}
\end{equation}

What happens if we send a plane wave with frequency $\omega_q$
to the DB? We will deal with the case of {\it one-channel}
scattering as for any $k \neq 0$ and any $q'$, $\omega_{q'} \neq
\omega_q + k\Omega_b$. This condition determines that all channels
with nonzero $k$ are 'closed', i.e. the spatial amplitudes
$e_{nk}$ are localized in space. Note here, that the frequencies
of transmitted and reflected waves are the same and coincide with
the the frequency of the incoming wave, since they all belong to
the only open channel with $k=0$.

For harmonic interaction potentials $W$ it was shown in
Ref.~\cite{tcsasf98} that
the momentum
\begin{equation}
J = -W''[0] \langle {\rm Im} \epsilon^*_n \epsilon_{n-1} \rangle
\label{momentum}
\end{equation}
is independent of the lattice site.
Here the averaging is meant with respect to time.
In a similar way it is
straightforward to show that the quantity
\begin{equation}
\tilde{J} = -\langle W''[\hat{x}_n(t)-\hat{x}_{n-1}(t)]
{\rm Im} \epsilon^*_n \epsilon_{n-1} \rangle
\label{tildemomentum}
\end{equation}
is independent of the lattice site. Since the breather solution
$\hat{x}_n(t)$ is spatially localized, at large distances from the
breather we find again that the momentum $J$ is independent of the
lattice site. Especially we find that it is the same for $n \rightarrow
\pm \infty$. Following Ref.~\cite{tcsasf98} we conclude that the one-channel scattering case is 
{\it elastic}, i.e.
the energy of an incoming wave equals the sum of the energies
of the outgoing (reflected
and transmitted) waves.

\begin{figure}[htb]
\vspace{20pt}
\psfrag{w}{$\omega_q$}
\psfrag{w-W}{$\omega_q-\Omega_b$}
\psfrag{w-2W}{$\omega_q-2\Omega_b$}
\psfrag{w-3W}{$\omega_q-3\Omega_b$}
\psfrag{w+W}{$\omega_q+\Omega_b$}
\psfrag{w+2W}{$\omega_q+2\Omega_b$}
\psfrag{w+3W}{$\omega_q+3\Omega_b$}
\centerline{\psfig{figure=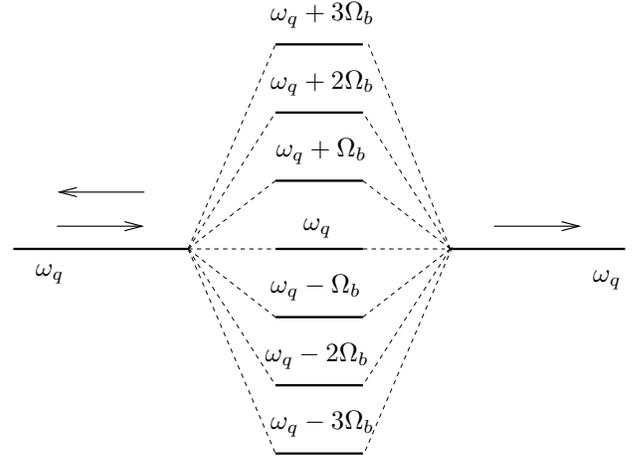,width=82mm,height=60mm}}
\vspace{2pt}
\caption{Schematic representation of the one-channel scattering of a wave by
a discrete breather. }
\label{fig2}
\end{figure}

Despite the fact that we will study the one-channel elastic
scattering, the scattering potential of DB is {\it time-periodic}.
The above mentioned 'closed' channels are active inside the
breather core, i. e. the frequency of linear waves in a finite
area around the breather center can change due to the interaction
with the DB (see Fig.\ref{fig2}).
%
Thus one
of the questions to be answered below is to identify the cases when
the well-known scattering by a {\it time-averaged (static)} potential
is not sufficient to describe the actual results of wave scattering
by discrete breathers. This implies that interference effects through
local interactions between the active closed channels may substantially
change the scattering results as compared to a time-averaged 
scattering potential.

Assuming that the breather is located around the central site $n=0$,
the one-channel scattering problem can be written as
\begin{eqnarray}\label{setup}
\epsilon_n(t)=A_I e^{-i(\omega_q t+qn)}+A_Re^{-i(\omega_q
t-qn)},~~n~<~0\nonumber\\
\epsilon_n(t)=A_Te^{-i(\omega_q t-qn)}~~, n~>~0
\label{setup2}
\end{eqnarray}
for
\begin{equation}
|n| \gg \sup [\xi_b(0),\xi_b(\omega_q)]
\label{largedistance}
\end{equation}
where $\xi_b(\omega)=\sup\xi(\omega + k\Omega_b)$ and
\begin{eqnarray}
  \begin{array}{l}
\sinh^2\xi(\nu)/2=\frac{V^{\prime\prime}[0]-\nu^2}{
4W^{\prime\prime}[0]}\;\;,\;\;
|\nu|<\omega_q(0)\\
\cosh^2\xi(\nu)/2=\frac{\nu^2-V^{\prime\prime}[0]}{
4W^{\prime\prime}[0]}\;\;,\;\; |\nu|>\omega_q(\pi)
  \end{array}
\end{eqnarray}
$\xi$ measures the characteristic inverse 
localization length of a closed channel
at frequency $\nu$ (note that for $\nu = k\Omega_b $ the localization
length is that of the breather itself,
see also \cite{tcsasf98}).
The incoming wave has amplitude $A_I$, and the reflected and transmitted
wave amplitudes are given by $A_R$ and $A_T$ respectively. The transmission
coefficient
$t_q=\left|A_T / A_I \right|^2$.

For further considerations we will use the notation of the
Bloch functions $\zeta_n(t)$ defined as
\begin{eqnarray}
\epsilon_n(t)=\zeta_n(t)e^{-i\omega_qt}.
\end{eqnarray}

In order to estimate the relative strength of closed channels
with $k \neq 0$ we
expand the time-periodic coefficients of (\ref{scat1}) in a Fourier series with respect to time:
\begin{eqnarray}
W^{\prime\prime}[\hat{x}_n(t)
-\hat{x}_{n-1}(t)] = \sum_{k=-\infty}^{\infty}
w_{n,k} {\rm e}^{{\rm i}k\Omega_b t} \label{fsp-w}\;,\\
V^{\prime\prime}[\hat{x}_n(t)] =
\sum_{k=-\infty}^{\infty}
v_{n,k}{\rm e}^{{\rm i}k\Omega_b t} \label{fsp-v}\;.
\end{eqnarray}
We consider first the case of a strongly localized optical breather 
located  at site $n=0$ with
$W(y)=\frac{c}{2}y^2$.
Taking into account a single closed channel with some value of $k$, inserting
(\ref{fsp-w}),(\ref{fsp-v}) and (\ref{Bloch2}) into (\ref{scat1}) and excluding
$e_{nk}$, the relative strength $s_k$ of the closed channel to the open one will
be given by
\begin{equation}
\label{strength}
s_k = \left| \frac{v_{0,k}^2}{(v_{0,0}-1)(v_{0,0}+2c(1-\eta_k)-(\omega_q+k \Omega_b)^2)}
\right| \;,
\end{equation}
where the relative amplitude 
\begin{equation}
\eta_k=\pm \frac{e_{1,k}}{e_{0,k}}={\rm e}^{-\xi(\omega_q+k \Omega_b)}
\label{relamp_opt}
\end{equation}
has positive sign for $\omega_q+k \Omega_b$ located inside the phonon gap
and negative sign otherwise.
For $s_k \ll 1$ we do not expect any significant contribution from the given
closed channel, while $s_k \geq 1$ indicate a strong influence of the closed
channel on the scattering process. Note that the 
expression $\sqrt{v_{00}+2c(1-\eta_k)}$ in the denominator of 
(\ref{strength}) is just a frequency $\omega_{L}$ of a {\it local phonon mode}
 of the time-averaged scattering potential. For spatially discrete 
systems these local phonon modes may be located 
inside or outside of the phonon gap. 
The denominator of (\ref{strength}) may vanish for certain wave numbers, 
which would imply a resonance-like enhancement of the closed channel 
contribution (for certain wavenumbers
$q$). As such a resonant enhancement of a closed channel amplitude acts as a
huge effective scattering potential to the open channel, for these cases we
expect a {\it resonant suppression of transmission}. Thus, qualitatively such a 
complete suppression of transmission (Fano-like resonance \cite{uf61,cskams98}) 
is explained by a resonant interaction of 
the propagating phonon with the specific local phonon mode. 
However in order to quantitatively 
analyze this effect  the renormalization of the value of $\omega_{L}$ due to 
all nonresonant processes, has to be taken into account. It will be done below 
for a particular case of optical breathers by making use of the Green function 
method. We obtain that although the renormalization of $\omega_{L}$ is rather 
small it may become important especially as the width of a phonon band is 
small, $W^{\prime\prime}~\ll~1$.

In a similar way we proceed for the estimation of the closed channel
contribution of acoustic breathers. We obtain 
the following expression for the relative strength $r_k$
\begin{equation}
\label{acousticstrength}
r_k = \left| \frac{w_{0,k}^2}{(w_{0,0})(w_{0,0}-(\omega_q+k \Omega_b)^2)}
\right|~~.
\end{equation}
Thus, we again find a 
resonant suppression of transmission when the presence of acoustic 
DB leads to a local {\it increase} in the nearest neighbor interaction and 
to an appearance of a corresponding local phonon mode. 
However, this case is much more involved as compared to the optical DB case, and
Eq. (\ref{acousticstrength}) may serve only as a qualitative tool to check whether a closed channel is 
strongly contributing to the transmission or not.
We will instead provide with a more quantitative analysis for the cases 
considered, based on the particular acoustic breather properties.

\subsection{Numerical scheme}

To compute numerically the transmission coefficient
we have developed a scheme which generalizes the one given
in  Ref. \cite{tcsasf98} (which relies on a spatial reflection symmetry
of the breather and thus of the scattering potential). At variance with Ref.
\cite{tcsasf98} our scheme
is capable of dealing with any (perhaps spatially nonsymmetric)
time-periodic scattering potential.

We look for solutions
of Eq.(\ref{scat1}) which correspond to zeroes of the operator
\begin{eqnarray}
\label{newton}
{\mathbf G}(\vec{\epsilon}(0),\dot{\vec{\epsilon}}(0))=
\left(\begin{array}{c}\vec{\epsilon}(0)\\
\dot{\vec{\epsilon}}(0)\end{array} \right)-e^{i\omega_q
T_b}\left(\begin{array}{c}\vec{\epsilon}(T_b)\\
\dot{\vec{\epsilon}}(T_b)\end{array}
\right)
\end{eqnarray}
on a lattice with $2N+1$ sites labeled $-N,(-N+1),...,-1,0,1,...(N-1),N$.
The incoming wave is fed from the left, and the transmitted wave
is leaving the system to the right. The boundary condition at the right end
is $\epsilon_{N+1}=e^{-i\omega_q t}$, which implies that the transmitted
wave will have amplitude 1. With a given boundary condition at the left
end $\epsilon_{-N-1}=(A+iB)e^{-i\omega_q t}$, where $A$ and $B$ are
real numbers, we may find the zeroes of (\ref{newton}) using
a standard Newton routine. Due to the linearity of the equations
of motion in $\epsilon$ an arbitrary initial guess and one Newton
step are needed to converge to the zeroes. In practice due to roundoff
errors an additional Newton step may be required.

However with
arbitrary $A$ and $B$ we will not realize the scattering case
(\ref{setup},\ref{setup2}) in general. This is due to the fact that
all extended Floquet states of an infinite system are twofold degenerated
because time reversal symmetry holds far from the breather center.
To succeed we add a second Newton procedure which uses $A$ and $B$
as free parameters such that the solution on site $N$ becomes
$\epsilon_N=e^{-iq-i\omega_qt}$, ensuring that we realize a single
transmitted traveling wave of amplitude one at the right end of the system.
After the Newton procedures are completed, the transmission coefficient
is then given by
\begin{eqnarray}
\label{numresult}
t_q=\frac{4\sin^2q}{|(A+iB)e^{-iq}-\zeta_{-N}|^2}.
\end{eqnarray}
While the Bloch functions $\zeta_n$ are in general time-dependent
close to the breather center, they will be time-independent
complex numbers at large distance from the breather.
We note that the computation operates at the machine precision,
and we obtain results which are size independent, i.e. with
the above described boundary conditions we emulate an infinite
system. The discrete breather solution itself has to be obtained
beforehand using standard numerical procedures \cite{tcsa97}.

The enumerator in (\ref{numresult})
vanishes at the extremal values of $\omega_q$, i.e. at
$q=0$ and $q=\pi$. If the denominator is finite at these values
of $q$, the transmission will also vanish. This is indeed the generic
case for a quadratic dependence of the spectrum $\omega_q$
on $q$ around these points. Exceptions are expected for
acoustic chains (see next Section).

Another peculiar point is that if upon changing some control parameter, e.g. the breather
frequency, a localized Floquet eigenstate attaches to or
disattaches from the extended Floquet spectrum, the transmission
coefficient will be exactly $t=1$ for the $q$-value which corresponds
to the edge of the spectrum $\omega_q$, i.e. $q=0$ or $q=\pi$
\cite{tcsasf98,swksk00}.

\subsection{Green function method for a time-periodic localized scattering potential}

\begin{figure}[htb]
\vspace{20pt}
\psfrag{n0}{\tiny $n_0$}
\psfrag{n1}{\tiny $n_1$}
\psfrag{n2}{\tiny $n_2$}
\psfrag{w}{\tiny $\omega_{\!q}$}
\psfrag{W}{\tiny $\Omega_{\!b}$}
\psfrag{2W}{\tiny $2\Omega_{\!b}$}
\psfrag{w-2W}{\tiny $\omega_{\!q}\!\!-\!\!2\Omega_{\!b}$}
\psfrag{w-4W}{\tiny $\omega_{\!q}\!\!-\!\!4\Omega_{\!b}$}
\psfrag{+}{\small $+$}
\psfrag{a}{a)}
\psfrag{b}{b)}
\psfrag{c}{c)}
\psfrag{d}{d)}
\centerline{\psfig{figure=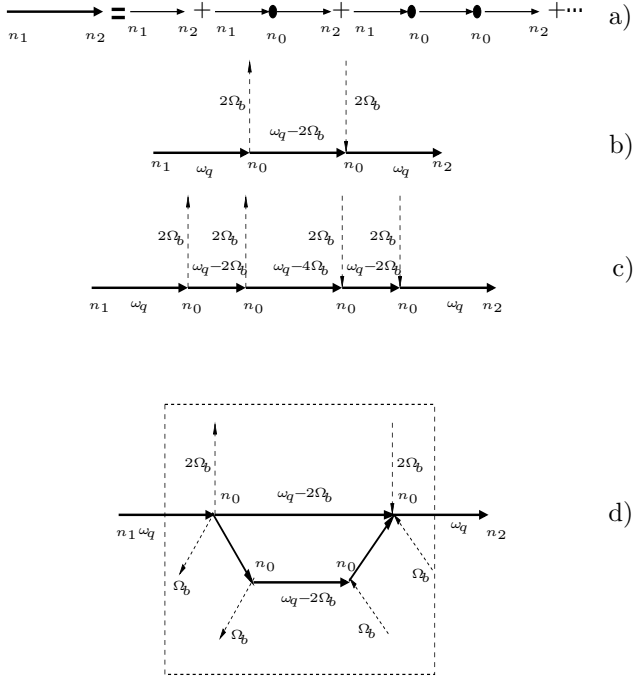,width=82mm,height=90mm}}
\vspace{2pt}
\caption{Typical diagrams describing the interaction of the propagating
phonons with the time-periodic DB scattering potential: a) scattering by 
a time-average (static) potential; b) resonant scattering process; c) 
a process leading to the renormalization of a phonon local mode frequency; d) 
various complex processes. Here, thin and thick
solid lines present the Green functions in the absence of the 
DB
and in the presence of the time-average part of the DB scattering
potential. The time-average part of the
DB scattering potential is shown by a black circle, and the dashed lines
show the absorption (emission) of phonons.}
\label{diagram}
\end{figure}

We will also analyze the wave propagation through DBs by making use of the Green function
technique \cite{ene90}. This method is especially convenient as the scattering potential is localized in
space.
To apply this method to a particular case when the presence of DBs leads to
the appearance of a localized time-dependent on-site scattering potential,
$V^{\prime \prime} [\hat x_n(t)]-1$,
we perform a time Fourier transformation of Eq. (\ref{scat1})
and obtain the equation for the Green function
$G_{\omega_q}(n_1,n_2)$ :
$$
G_{\omega_q}(n_1,n_2)~=~G^0_{\omega_q}(n_1,n_2)-
$$
\begin{equation} \label{GreenFunct}
-\sum_m G^0_{\omega_q}(n_1,m)
\int d\Omega U_\Omega (m) G_{\omega_q+\Omega}(m,n_2)~~,
\end{equation}
where $G^0_{\omega_q}(n_1,n_2)$ is the Green function of
the {\it linear} equations of motion in the absence of the DB.
The Fourier transform of the localized scattering potential
$U_\Omega (m) ~=~\int dt e^{i\Omega t} (V^{\prime \prime} [\hat x_m(t)]-1)$
is determined by the properties of the DB solution. Because the DBs are periodic
solutions in time the potential $U_\Omega (m)$ contains just the harmonics of the
breather frequency, $\Omega~=~k \Omega_b$. Moreover, as we will see later in
the considered cases
we can take into account the harmonics with small values of $k~=0~,\pm 1, ~\pm 2 $
only.
Thus, the calculation of the Green function $G_{\omega_q}(n_1,n_2)$ can be 
represented in a diagrammatic form where terms which correspond to active closed channels 
describe the local (virtual) absorption and emission
of phonons by the propagating phonon  in the presence of the DB
(some of the typical diagrams are shown in Fig.\ref{diagram}).
Moreover, for
one-channel scattering the energy conservation law of
absorbed and emitted phonons has to hold.
In order to obtain the transmission coefficient $t_q$ we need also an expression for the
Green function $G^0_{\omega}(n_1,n_2)$ which reads
\begin{equation} \label{GreenFunctSol}
G^0_{\omega}(n_1,n_2)~=~-\int \frac{dq}{2\pi} \frac{e^{iq (n_1-n_2)}}{\omega^2-\omega_q^2}~~.
\end{equation}
The Green function of the full problem (for one-channel scattering)
has a similar form for large distances from the breather center
($n_1 \rightarrow -\infty$ and $n_2 \rightarrow +\infty$):
\begin{equation} \label{GreenFunctSolGen}
G_{\omega_q}(n_1,n_2)~=~-i D_q\frac{e^{iq |n_1-n_2|}}{d(\omega_q^2)/dq}~~,
\end{equation}
where $t_q~=~|D_q|^2$.

As an example for a static scattering potential which is strongly
localized on a single site ($n=0$) we obtain
\begin{equation} \label{GreenFunctSolStat}
G_{\omega_q}(n_1,n_2)~=~\frac{G_{\omega_q}^0(n_1,n_2)}{1+\beta_0 G_{\omega_q}^0(0,0)}~~,
\end{equation}
where $\beta_0$ is the strength of the static potential.
The transmission coefficient $\tilde t_q$ in this case is given by
\begin{equation} \label{TransCoeffStat}
\tilde t_q  ~=~\frac{(d(\omega_q^2)/dq )^2}{\beta_0^2+(d(\omega_q^2)/dq)^2}~~.
\end{equation}
We will use the Green function method in Section V
where the wave scattering by optical DBs (see Fig. 1c) will be presented. We show how most
important diagrams can be taken into account in the case of a {\it time-periodic localized}
scattering potential.
%

\subsection{Time-averaged scattering potential}
Since we are interested in understanding the importance of the time dependence
of the scattering potential, we will also compare the numerical results
with those obtained by time-averaging the DB scattering potential.
This time-averaging can be found numerically beforehand and Eq. (\ref{scat1})
becomes
\begin{eqnarray}\label{scat2}
\ddot{\tilde{\epsilon}}_n&=&
-w_{n,0}
(\tilde{\epsilon}_n-\tilde{\epsilon}_{n-1})+\nonumber\\
&&w_{n+1,0}
(\tilde{\epsilon}_{n+1}-\tilde{\epsilon}_n)-
v_{n,0} \tilde{\epsilon}_n.
\end{eqnarray}
Because in this case all inhomogeneities are time-independent we can use
the standard scattering matrix method.
With $\tilde{\epsilon}_n(t)=\tilde{\zeta}_ne^{-i\omega_qt}$
Eq.(\ref{scat2}) is rewritten as
\begin{eqnarray}
\left(\begin{array}{l}\tilde{\zeta}_{n+1}\\ \tilde{\zeta}_n \end{array}
\right)={\mathbf M}_n\left(\begin{array}{l}\tilde{\zeta}_{n}\\
\tilde{\zeta}_{n-1}
\end{array} \right),
\end{eqnarray}
with
\begin{eqnarray}\label{matrix}
{\mathbf M}_n=\left(\begin{array}{cc}1+\frac{E_n+c_{n,n-1}-\omega_q^2}{c_{n+1,n}}&-\frac{c_{n,n-1}}{c_{n+1,n}}\\1&0
\end{array}\right)
\end{eqnarray}
where $E_n=v_{n,0}$ and
$c_{n,n-1}=w_{n,0}$.
It follows
\begin{eqnarray}
\left(\begin{array}{l}\tilde{\zeta}_{N+1}\\
\tilde{\zeta}_N \end{array}
\right)={\mathbf M}\left(\begin{array}{l}\tilde{\zeta}_{-N}\\
\tilde{\zeta}_{-N-1}
\end{array} \right),
\end{eqnarray}
with
\begin{eqnarray}\label{M}
\mathbf{M}=\prod\limits^{-N}_{i=N}\mathbf{M}_i.
\end{eqnarray}

The expression for the transmission coefficient $\tilde{t}_q$ for
the time averaged scattering potential is then given by
\begin{eqnarray}\label{trans}
\tilde{t}_q=\frac{4\sin^2q}{|M_{11}(q)e^{-iq}+
M_{12}(q)-M_{21}(q)-M_{22}(q)e^{iq}|^2}
\end{eqnarray}
where $M_{ij}$ are the four matrix elements of the $2\times 2$
matrix $\mathbf{M}$.

\section{Scattering by acoustic breathers}

In this Section we study the wave scattering by so-called
{\it acoustic breathers}. The corresponding systems are characterized
by a gapless spectrum $\omega_q$ of propagating linear waves, and by a conservation of
total mechanical momentum.

The generic choice for the potentials in the Eq. (\ref{Mot}) is
$V(y)\equiv0$ and
$W(y)=\frac{1}{2}y^2+\frac{\beta}{3}y^3+\frac{1}{4}y^4$. This choice leads to
the well known Fermi-Pasta-Ulam system \cite{jlmsa96}.
We will first consider the case $\beta=0$ which implies
the presence of a parity in the interaction potential $W$ and
comment on the influence of parity violation for $\beta \neq 0$ later on.
The dispersion relation of phonons is given by
\begin{eqnarray}
|\omega_q|=2\sin\frac{q}{2}
\end{eqnarray}

\begin{figure}[htb]
\vspace{20pt} \psfrag{x}{$\hat{x}_n(t=0)$}
\centerline{\psfig{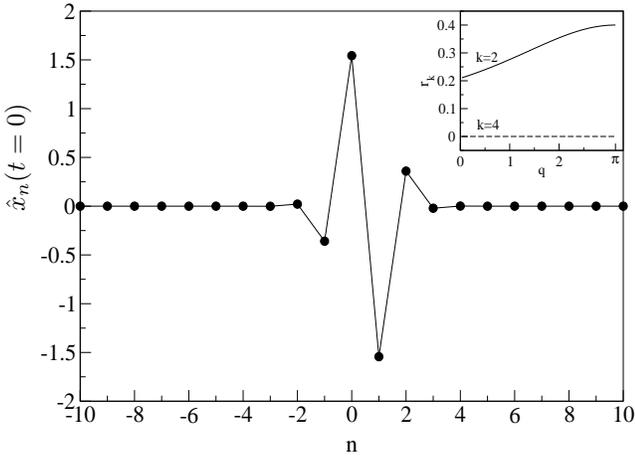}}
\vspace{2pt} \caption{Displacements of an
acoustic breather with zero velocities at $t=0$ and
$\Omega_b=4.5$.
\\
Inset: Relative strength $r_k$ for the second and fourth closed
channels versus $q$.} \label{fpu_profile}
\end{figure}

\begin{figure}[htb]
\vspace{20pt}
\psfrag{w}{$|w_{n,k}|$}
\centerline{\psfig{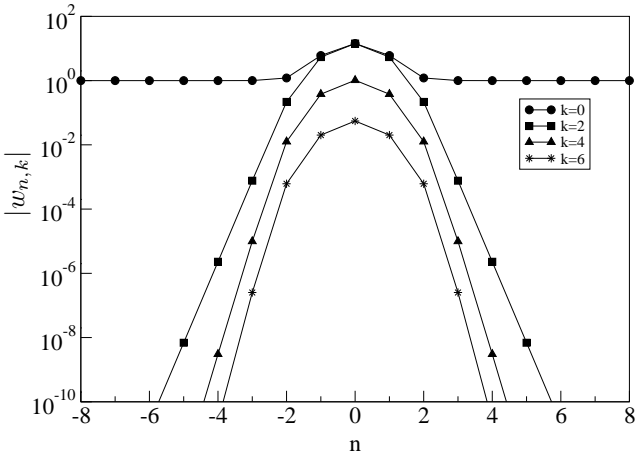}}
\vspace{15pt}
\caption{Fourier components $w_{nk}$ for different $k$ versus $n$ for the
breather in Fig.\ref{fpu_profile}.}
\label{fpu_harm}
\end{figure}

A breather solution with frequency $\Omega_b=4.5$ is shown in Fig.\ref{fpu_profile}.
%
The corresponding Fourier components of the scattering potential are plotted
in Fig.\ref{fpu_harm}.
%
Acoustic breathers can be found
with frequencies above the phonon band $|\Omega_b| > \max{|\omega_q|}$.
However, here we consider  $|\Omega_b| > 2 \max{|\omega_q|}$ which implies
$|\Omega_b| > 4$. This condition is needed to realize the one-channel scattering case
of linear waves. For $\beta=0$ the ac part of the 
interaction potential $W$ has even 
harmonics only,  and the frequency of a propagating phonon in a closed channel 
$2\Omega_b \pm \omega_q$ can not match the frequency of a local mode
of the time-averaged scattering potential.  
Thus the closed channels play no important role in the scattering process.
As a consequence {\it the wave scattering by
acoustic breathers is practically
identical with the scattering by the time-averaged
potential}. Indeed numerical computations do not show any relevant
difference between the linear wave propagation in the presence of time-dependent
and the static (time-averaged)
potentials. Yet there is a number of interesting
features of the scattering which deserve to be exploited.

For $V=0$ the lattice conserves the total mechanical momentum
$P=\sum_n \dot{x}_n$ in addition to the energy. Without loss
of generality we will choose $P=0$ here. From this conservation
law it follows that the total mechanical momentum of
the linearized problem $\Pi = \sum_n \dot{\epsilon}_n$ is also
conserved. This implies  that for any solution
$\epsilon_n(t)$ the shift $\epsilon(t)+C$ is also a solution.
In particular $\epsilon_n = C$ is a solution, which corresponds
to a wave with $q=0$ and for which we have $t_{q=0}=1$. Thus we conclude
that the transmission of waves by a breather for acoustic systems at $q=0$ is always
perfect, no reflection occurs. This is in sharp contrast to
systems without conservation of total mechanical momentum (see below).

\begin{figure}[htb]
\vspace{20pt}
\centerline{\psfig{figure=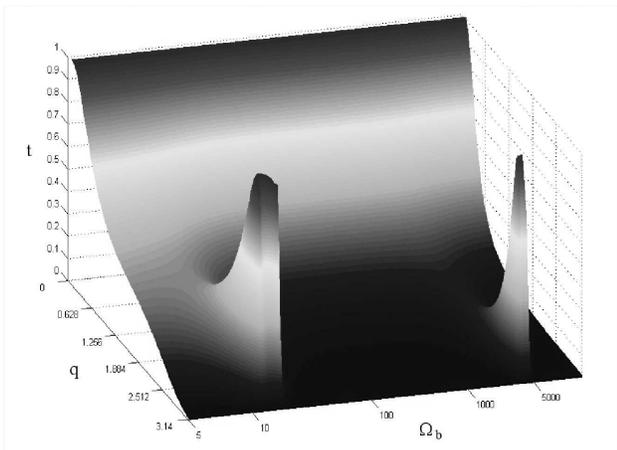,width=82mm,height=60mm}}
\vspace{2pt}
\caption{The dependence of the transmission coefficient on $q$ and $\Omega_b$.}
\label{fig4}
\end{figure}

\begin{figure}[htb]
\vspace{20pt}
\psfrag{W}{$\Omega_b$}
\centerline{\psfig{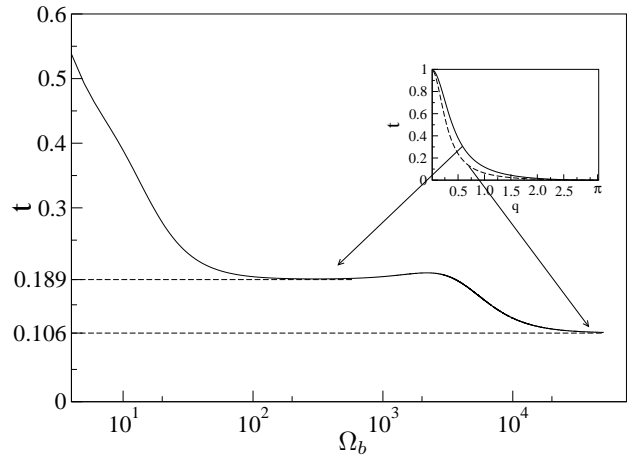}}
\vspace{2pt}
\caption{The dependence of the transmission coefficient on $\Omega_b$
for a particular value of $q=\frac{\pi}{4}$.
The dashed lines show the predicted plateau heights (see text).
The inset shows the dependence of the transmission on $q$
for the two observed plateau regions.}
\label{fig5}
\end{figure}

The peculiar dependence of the transmission coefficient $t_q$ on the wave number
$q$ and the breather
frequency $\Omega_b$ is shown in Fig.\ref{fig4}. The breather frequency is
varying over nearly three decades. We indeed observe perfect transmission
at $q=0$, and zero transmission at $q=\pi$. However we also find
that in the studied breather frequency range two rather narrow peaks 
around $q=\pi$
appear with perfect transmission. These structures are due to the detachment
of localized Floquet eigenvectors from the continuum of extended Floquet eigenstates.
%
A surprising result is shown in Fig.\ref{fig5}. We plot
the transmission $t_q$ at $q=\pi/4$ as a function of the breather frequency $\Omega_b$.
We obtain plateaus and crossovers at certain breather frequencies.
The crossover positions clearly correlate with the appearance of
localized Floquet states, which are traced
through the perfect transmission close to $q=\pi$ in Fig.\ref{fig4}.
The dependence of the
transmission on $q$ for the two observed plateaus is shown in the
inset of Fig.\ref{fig5}. The plateaus range over several decades
in $\Omega_b$, and the $q$-dependence of $t_q$ is rather similar on different plateaus.

To understand this feature we remind that large breather
frequencies imply that the breather energy and its amplitude in
the breather center are large as well. Thus we may neglect the
harmonic part of the interaction potential $W$ inside the breather
core. The resulting amplitude distribution of the breather core is
characterized by a superexponential decay $\hat{x}_n(t) = A_n
G(t)$, $A_n \sim A_{n-1}^3$, where $G(t)$ is an oscillatory master
function \cite{sf94,bdmesfgpt01}. Due to the symmetry of the
breather solution the spatial profile is described by
$A_{-n} = -A_{n+1}\;,\;n\leq 0$.
The amplitudes $A_n$
have been computed: $A_1=1$, $A_2 ~\simeq~-0.166$, $A_3
~\simeq~ 4.796 ~10^{-4}$, $A_4 ~\simeq~ -1.15 ~10^{-11}$, and so
on \cite{jlmsa96}. The overall amplitude of the oscillations
$\hat{x}_n(t)$ is tuned by the amplitude of the master function
$G(t)$.

The scattering is essentially described by the time-averaged
scattering potential. This potential corresponds to
a large {\it increase} in the nearest neighbour coupling terms.
In the breather center we find
\begin{equation}
\langle W''[\hat{x}_1(t)-\hat{x}_0(t)] \rangle =1 + 3\langle
(\hat{x}_1(t)-\hat{x}_0(t))^2 \rangle
\end{equation}
with
\begin{equation}
3\langle
(\hat{x}_1(t)-\hat{x}_0(t))^2 \rangle \equiv X^2 \approx
\frac{2}{3\pi}\Omega_b^2
\end{equation}
(the details of the calculation are presented in Appendix A).
Due to the large value of $X$ for large $\Omega_b$
it becomes evident that time-dependent corrections to the scattering
potential are negligible. By making use of the distribution of oscillation amplitudes
in the lattice
we then find
\begin{equation}
3\langle
(\hat{x}_n(t)-\hat{x}_{n-1}(t))^2 \rangle
\approx A_n^2 X^2.
\end{equation}
The scattering on a plateau in Fig.\ref{fig5} is due to a finite
number of matrices $\mathbf M_n$ which should be taken into account in Eq. (\ref{M}).
Qualitatively the crossover threshold
can be defined as
\begin{equation}
A_n^2 X^2 \simeq 1~~.
\end{equation}
 Thus we obtain the two crossover frequencies
$\Omega_{b1}=13.1$ and $\Omega_{b2}=4527$, which match with
the observed crossover positions. For
$4<\Omega_b<\Omega_{b1}$ we need
only 4 matrices $\mathbf M_n$, for
$\Omega_{b1}<\Omega_b<\Omega_{b2}$, 6 matrices $\mathbf M_n$ and so on.
We also computed the transmission at $q=\pi/4$ using the corresponding
reduced set of relevant matrices. The obtained limiting values of $t_q$
are
shown as a dashed line in Fig.\ref{fig5}. We observe very good agreement
between the predicted plateau heights and the actual data obtained in
direct numerical simulations.

%

%

With the above said it becomes also transparent, why we observe
detachment of localized Floquet states in the crossover region.
The acoustic breather presents an {\it effective potential well}
for the propagating phonons.
The width of this well increases with the breather frequency, and the
number of possible localized Floquet states also increases. Such a
periodic appearance of perfect transmission
is similar to the quantum mechanical scattering by a potential well
in the presence of {\it quasi-discrete} levels \cite{MorFish}.

\begin{figure}[htb]
\vspace{20pt}
\centerline{\psfig{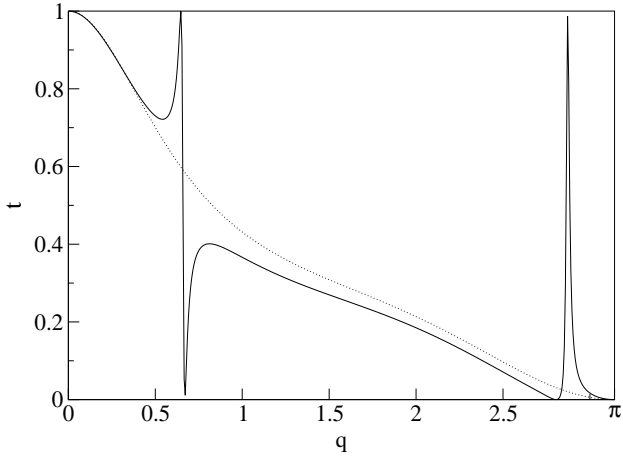}}
\vspace{2pt}
\caption{The dependence of the transmission coefficient on $q$
for an acoustic breather with $\Omega_b=4.5$ and $\beta=1$.
The dashed line is the result for the time-averaged scattering potential.
Two transmission zeros are observed at corresponding values of $q$.}
\label{figfpu34}
\end{figure}

Finally we checked the influence of $\beta \neq 0$.
With respect to the breather the presence of cubic terms in the
interaction potential leads to the generation of a kink-shaped
dc lattice distortion. The corresponding scattering potential
becomes asymmetric in space and odd harmonics in the ac scattering 
potential appear. This immediately leads to the possibility of matching 
between the  frequency of a propagating phonon in a closed channel 
$\Omega_b-\omega_q$ and several local modes of the time-averaged scattering potential. 
Thus, a resonant suppression of a transmission appears in this case. 
Indeed, we observe this effect by direct numerical simulations as shown
in Fig.\ref{figfpu34}, where two transmission zeros are found. We will discuss
this effect in more detail for the case of optical breathers below. 
%

\section{Scattering by acoustic rotobreathers}

\begin{figure}[htb]
\vspace{20pt} \psfrag{x}{$\hat{x}_n(t=0), \dot{\hat{x}}_n(t=0)$}
\centerline{\psfig{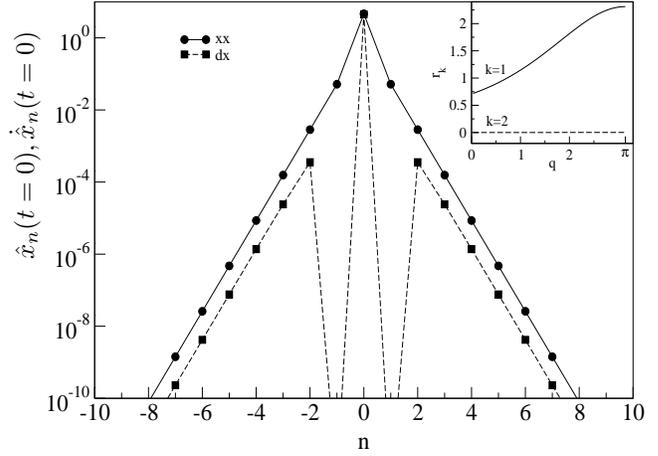}}
\vspace{2pt} \caption{Displacements and
velocities of an acoustic rotobreather at $t=0$ with
$\Omega_b=4.5$. 
\\
Inset: Relative strength $r_k$ for the first and second closed
channels versus $q$.
} \label{rotofpu_profile}
\end{figure}

\begin{figure}[htb]
\vspace{20pt}
\psfrag{w}{$|w_{n,k}|$}
\centerline{\psfig{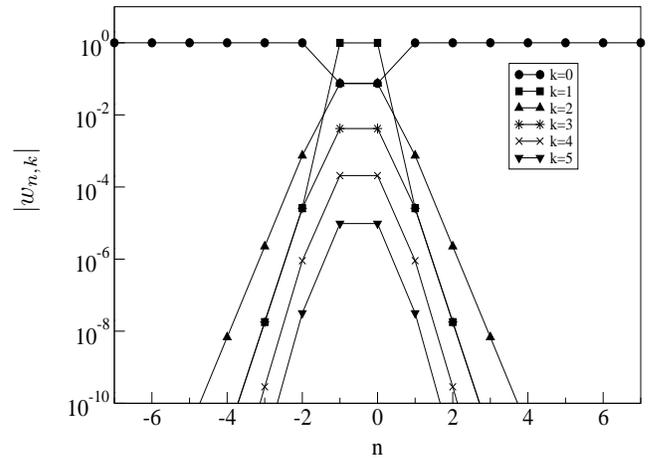}}
\vspace{2pt}
\caption{Fourier components $w_{nk}$ for different $k$ versus $n$ for the
rotobreather in Fig.\ref{rotofpu_profile}.}
\label{rotofpu_harm}
\end{figure}

\begin{figure}[htb]
\vspace{20pt}
\psfrag{q}{q}
\centerline{\psfig{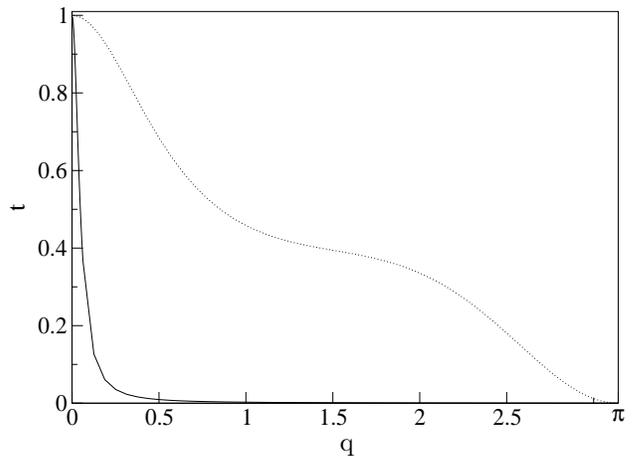}}
\vspace{2pt}
\caption{The dependence of the transmission coefficient on $q$ for $\Omega_b=4.5$.
Data for acoustic rotobreather and acoustic breather are shown
correspondingly by solid and dashed lines.}
\label{fig6}
\end{figure}

Remarkable differences to the results of the preceding
Section are obtained if the interaction potential $W$ is chosen to
be a periodic one (the potential $V$ is still zero in this Section):
\begin{equation}
W(y)=1-\cos(y).
\end{equation}
With such an interaction potential the nonlinear chain allows for the existence of
{\it rotobreathers} (cf. Fig.1b).
In the simplest case a rotobreather consists
of one particle being in a rotating (whirling) state, while all
others particles with spatially decaying amplitudes from the center of DB:
\begin{eqnarray}
\hat{x}_0(t+T_b)=\hat{x}_0(t)+2\pi \nonumber, \\
\hat{x}_{n \neq 0}(t+T_b) = \hat{x}_{n \neq 0}(t), \\
\hat{x}_{|n| \rightarrow \infty} \rightarrow 0 \nonumber.
\end{eqnarray}
To excite such a rotobreather the central particle needs to overcome
the potential barrier generated by its two nearest neighbors.
The rotobreather energy is thus bounded from below by $E_b > 4$.
At sufficiently large energies $E_b \gg 4$ the central particle
will perform nearly free rotations $\hat{x}_0(t) = \Omega_b t + \delta(t)$
with $\delta(t+T_b)=\delta(t)$ being a small correction.
The time-averaged off-diagonal hopping terms between site $n=0$ and
$n=\pm 1$ are then given by
\begin{equation}
\label{weaklink}
\langle \cos (\hat{x}_0(t) - \hat{x}_1(t)) \rangle \approx
- \langle \sin (\Omega_b t) (\delta(t) - \hat{x}_1(t) ) \rangle
\end{equation}
where $\hat{x}_1(t)$ is also a small function. Thus the time-averaged scattering potential
of a rotobreather presents a huge {\it barrier} that cuts the chain into nearly
noninteracting parts, contrary
to the previous case of an acoustic breather, where the acoustic breather potential
corresponds to a potential well. Therefore it becomes extremely difficult
for waves to penetrate across
the rotobreather.
The rotobreather solution with $\Omega_b=4.5$ is shown in Fig.\ref{rotofpu_profile}.
%
The corresponding Fourier components of the scattering potential
are shown in Fig.\ref{rotofpu_harm}.
In Fig.\ref{fig6} we show the $q$-dependence
of the transmission for a rotobreather with $\Omega_b=4.5$ and
compare it to the corresponding curve of an acoustic breather from
the preceding section. We observe a dramatic decrease of the
transmission at all $q$-values for the rotobreather, in agreement
with the above analysis.

\begin{figure}[htb]
\vspace{20pt}
\psfrag{W}{$\Omega_b$}
\centerline{\psfig{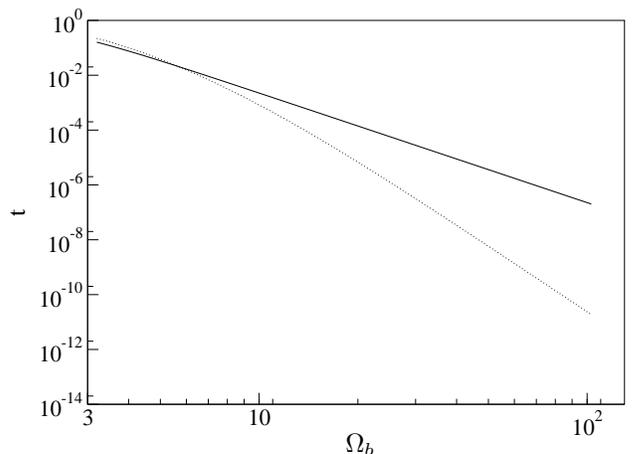}}
\vspace{2pt}
\caption{The dependence of the transmission coefficient on the frequency
of the breather $\Omega_b$ at fixed $q=0.21$.
Scattering by "real" rotobreather and a static part of DB are
shown correspondingly by solid  and dashed lines.}
\label{fig7}
\end{figure}

To decide whether the time-dependence of the rotobreather scattering
potential is important or not, we first make
a more precise computation for (\ref{weaklink})
(details of the calculations are presented in Appendix B)
\begin{equation}
\label{weaklink2}
\langle \cos (\hat{x}_0(t) - \hat{x}_1(t)) \rangle  \approx
 -\frac{3}{2\Omega_b^2}.
\end{equation}
The time-averaged scattering potential for large $\Omega_b$ corresponds
to two neighboring weak links of strength (\ref{weaklink2})
inserted in a linear acoustic chain. Three matrices $\mathbf M_n$ are
enough to compute the transmission. Two of them involve matrix elements
which are proportional to $\Omega_b^2$. The elements of the product
will thus contain terms proportional to $\Omega_b^4$, and according to
(\ref{trans}) the result is
\begin{equation}
\tilde{t}_q \sim \Omega_b^{-8}.
\end{equation}
This is precisely what we also find from a numerical evaluation
of the transmission for
the time-averaged scattering potential in Fig.\ref{fig7}.
However, although the transmission coefficient $t_q$
for the full time-dependent scattering potential also drastically decreases with $q$ and
$\Omega_b$, the dependence is weaker than for the case of time-average rotobreather potential.
It scales as
\begin{equation}
t_q \sim \Omega_b^{-4}
\label{truescaling}
\end{equation}
(see also Fig.\ref{fig7}).
The reason is that besides a weak static link the rotobreather scattering
potential has an ac term at frequency $\Omega_b$ of amplitude one (see
Fig.\ref{rotofpu_harm}).
Thus an alternative route for the wave is to approach the rotobreather,
to be excited into the first closed channel, to pass the breather and
to relax back into the open channel. The corresponding scattering
process of "virtual" absorption and emission of phonons from the rotobreather
 can be also represented by three matrices as it happens
for the dc analysis. However now instead of two weak links we have links
of order one with a frequency change at site $n=0$ from
$\omega_q$ to $\omega_q+\Omega_b$. This occurs in exactly one
of the three matrices. Consequently the product matrix will contain
elements proportional to $\Omega_b^2$ and the transmission will scale as
(\ref{truescaling}).

%
%

We conclude this section by stressing that {\it the wave scattering
by an acoustic rotobreather is essentially relying on the time-dependence
of the scattering potential}. The rotobreather effectively cuts the
chain in weakly interacting parts and thus hinders waves from propagation
in a very strong way.

\section{Optical breathers}

In this Section
we consider systems with a nonvanishing
on-site potential $V[x_n]\not=0$ (see Fig. 1c).
The difference of such systems to acoustic models
is the existence of a gap in the spectrum of phonons
$|\omega_{q=0}|=V^{\prime\prime}[0]$.
As a consequence the total mechanical momentum is not conserved,
and the transmission coefficient now vanishes not only at $q=\pi$
but also at $q=0$. An exception is the case when a localized Floquet eigenstate
bifurcates from the corresponding band edge for some special parameters
\cite{tcsasf98,swksk00,skcbrsm97}.
Because of the presence of a gap in the plane wave spectrum there
are now two different cases of interest - the breather frequency being
located outside the spectrum $|\Omega_b| > \max{| \omega_q |}$ or inside
the gap $|\Omega_b| < \min{|\omega_q |}$.

\subsection{The case $|\Omega_b| > \max{| \omega_q |}$.}

Here we choose
$V(y)=\frac{1}{2}y^2+\frac{1}{3}y^3+\frac{1}{4}y^4$ and
$W(y)=\frac{c}{2}y^2$.

\begin{figure}[htb]
\vspace{20pt} \psfrag{x}{$\hat{x}_n(t=0)$}
\centerline{\psfig{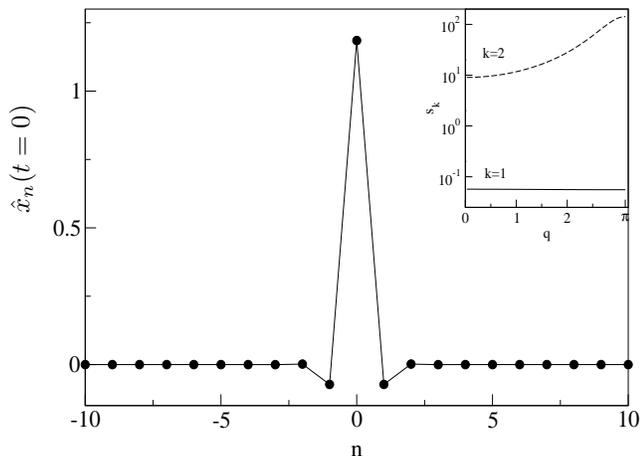}}
\vspace{2pt} \caption{ The initial displacements of an optical
breather with $\Omega_b=1.5$ and $c=0.05$ (velocities are zero).
\\
Inset: $s_{k=1,2}$ versus $q$.
}
\label{x34_profile}
\end{figure}

\begin{figure}[htb]
\vspace{20pt}
\psfrag{v}{$|v_{n,k}|$}
\centerline{\psfig{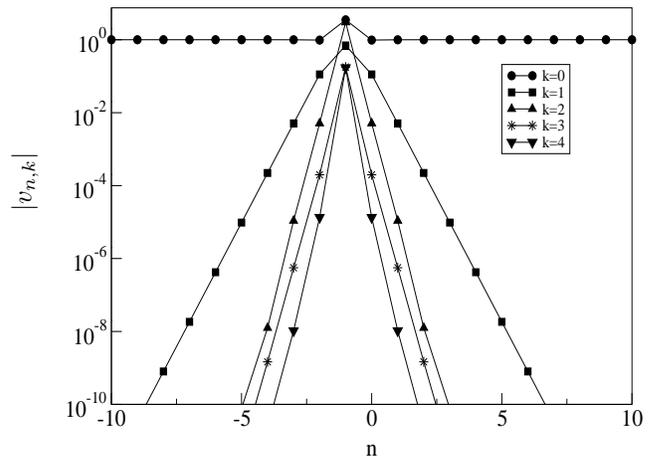}}
\vspace{2pt}
\caption{ Fourier components $v_{nk}$ for different $k$ versus $n$ for the
breather in Fig.\ref{x34_profile}.}
\label{x34_harm}
\end{figure}

In this case the spectrum of phonons is
\begin{equation}
\omega_q^2~=~1+4c\sin^2(\frac{q}{2})~~.
\end{equation}
The breather profile for $\Omega_b=1.5$ and $c=0.05$ is shown in Fig.\ref{x34_profile}.
The corresponding Fourier components of the scattering potential are plotted
in Fig.\ref{x34_harm}.
We note that the DB scattering potential perturbs the diagonal terms and
in this particular case ($\Omega_b > \omega_q$)
presents a {\it barrier} for propagating phonons.
For large breather frequencies
the breather is strongly localized, i.e. essentially only one central
oscillator is excited. The time-averaged scattering potential
then becomes a single diagonal defect with a large strength $\beta_0~\simeq~\Omega_b^2$.
It is straightforward to observe that the transmission coefficient
will thus scale as (see Eq.(\ref{TransCoeffStat}))
\begin{equation}\label{Scale}
\tilde{t}_q \sim \Omega_b^{-4}.
\end{equation}
Due to the fact that the transmission vanishes exactly for both
$q=0$ and $q=\pi$ we conclude that for large breather frequencies
transmission is suppressed in general.

\begin{figure}[htb]
\vspace{20pt} 
\psfrag{k}{\small $q$}
\psfrag{q}{$q$}
\centerline{\psfig{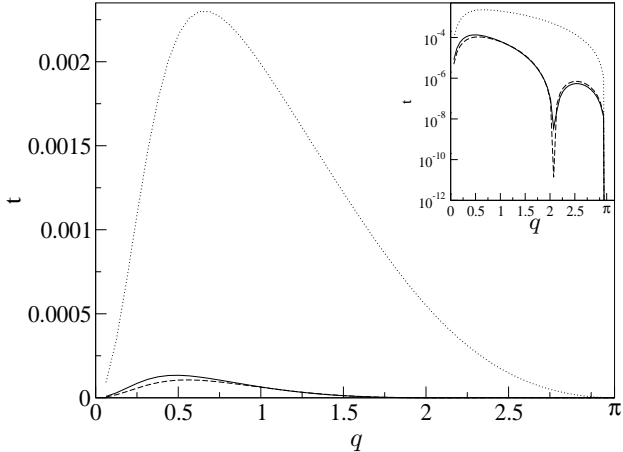}}
\vspace{2pt} \caption{The dependence of the transmission
coefficient $t_q$ on the wave number $q$. The optical breather
frequency is $\Omega_b=1.5$ and the coupling $c=0.05$. The results
are shown for the time-averaged DB scattering potential (dotted
line); the time averaged part and the second harmonic of the DB
scattering potential (dashed line); the full DB scattering
potential (solid line).
\\
Inset: same with $t_q$ on a logarithmic scale. Note the resonant
suppression around $q=2.1$.} \label{fig10}
\end{figure}

However, in the case of an optical DB
the time-dependent part of the DB scattering potential
(more precisely its second harmonic, see Fig. \ref{x34_harm} ) is also large
as $\beta_2~\simeq~\Omega_b^2$ for large values
of the breather frequency. Thus, the time-dependent
part of the DB potential can be rather important. 
Indeed, for the breather from Fig.\ref{x34_profile} the obtained transmission as
a function of $q$ shows that $\tilde{t}_q$ (see Fig.\ref{fig10}, dashed line) is at
least one order of magnitude larger than $t_q$ (see Fig.\ref{fig10}, solid line). In addition $t_q$ shows a resonant
minimum around $q=2.1$ (see inset). Let us use the estimation (\ref{strength}) 
for $k=1,2$. The parameters are $v_{0,0}=3.5\;,\;v_{0,1}=0.69\;,\;v_{0,2}=3$.
First we find that $s_1 \approx 0.05$ for all $q$ 
implying that the $k=1$ closed channel does not participate in the transmission process. At the same time 
$s_2 > 10$ for all $q$ and thus, the $k=2$ closed channel strongly participates in the transmission process. 
It is because the frequency of the propagating phonon $2\Omega_b-\omega_q~\simeq~1.92$ is close to 
a localized eigenmode of the time-averaged scattering potential $\omega_{L}~=~1.9$. 
Notice here that the frequency of this local 
mode is above the phonon band. 
Thus we interprete the suppression of transmission as a strong coupling between
the propagating wave and a particular localized mode of the time-averaged scattering potential,
mediated by the ac terms of the scattering potential (the $k=2$ channel in this case).
However, the simple estimation based on Eq. (\ref{strength}) does not show the resonant suppression of 
transmission for a particular value of $q$ and therefore, in order to quantitatively describe this effect 
we next apply the Green function method.

%

\begin{figure}[htb]
\vspace{20pt} \psfrag{W}{$\Omega_b$}
\centerline{\psfig{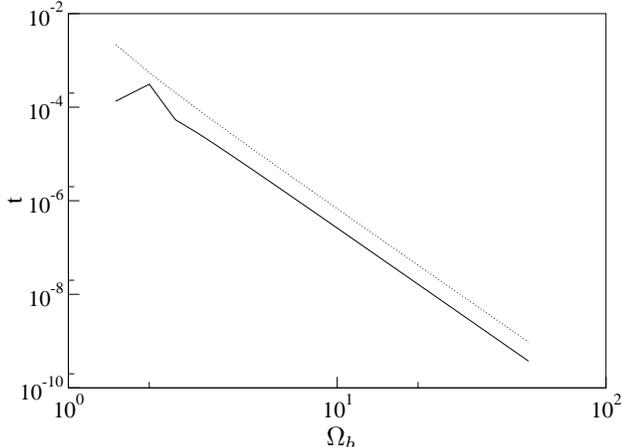}}
\vspace{2pt} \caption{The dependence of the transmission
coefficient $t_q$ on the optical breather frequency $\Omega_b$ for
a particular value of wave number $q=0.5$. The dashed line
corresponds to the scattering by a time-averaged DB scattering
potential.} \label{fig9}
\end{figure}

By making use of the estimation (\ref{strength}) we take into account in (\ref{GreenFunct}) the
terms with $k=0$ and $k=\pm 2$, i. e. the scattering on the static potential
and the phonon interaction with the second harmonic of the frequency
$2 \Omega_b $.
Moreover, we use the fact that the DB scattering potential is a strongly
localized one, and obtain (the corresponding diagram is shown in Fig. 3b)
$$
G_{\omega_q}(n_1,n_2)~=~\tilde G_{\omega_q}(n_1,n_2)+
$$
\begin{equation}\label{GrFunctOpt1}
+\beta_2^2 \tilde G_{\omega_q}(n_1,0)\tilde G_{-\Omega+\omega_q}^{(rn)}(0,0)
 G_{\omega_q}(0,n_2)~~,
\end{equation}
where $G_{\omega_q}(n_1,n_2)$ is determined by
$\tilde G_{\omega}(n_1,n_2)$ that is the Green function of propagating
phonons in the presence of time-average DB potential (the corresponding diagram is shown in Fig. 3a):
\begin{equation}\label{GrFunctOpt0ch}
\tilde G_{\omega}(n_1,n_2)~
=~G^0_{\omega}(n_1,n_2)-\beta_0 G^0_{\omega}(n_1,0)
\tilde G_{\omega}(0,n_2)~~ .
\end{equation}
Moreover, the central part of Eq. (\ref{GrFunctOpt1}), namely $\tilde G_{-\Omega+\omega_q}^{(rn)}(0,0)$, has a resonant 
form:
\begin{equation}\label{GrFunctRN}
\tilde G_{-\Omega+\omega_q}^{(rn)}(0,0)~
=~\frac{1}{\omega_L^2-(\Omega-\omega_q)^2}~~,
\end{equation}
where the local phonon mode frequency $\omega_L$ is mostly determined  
by a time-average scattering potential. 
However, in the case as the phonon band is narrow ($c~\ll~1$) the renormalization of $\omega_L$ due to 
the ac nonresonant processes has to be taken into account (the corresponding diagram is shown in Fig. 3c).
The Eqs. (\ref{GrFunctOpt1}) and (\ref{GrFunctOpt0ch}) can be solved and we obtain
\begin{equation}\label{GrFunctOpt1Sol}
G_{\omega_q}(0,n_2)~=~
\frac{G^0_{\omega_q}(0,n_2)}{1+
(\beta_0-
\beta_2^2\tilde G_{-\Omega+\omega_q}^{(rn)}(0,0))
G^0_{\omega_q}(0,0)}~~.
\end{equation}
Thus, we arrive at the expression (\ref{TransCoeffStat}) for the transmission
coefficient $t_q$ but with the renormalized wave number dependent
parameter $\beta$
\begin{equation}\label{betaRen}
\beta~=~\beta_0-\frac{\beta_2^2}{\omega_L^2-(2\Omega_b-\omega_q)^2}.
\end{equation}
As the breather frequency is large ($\Omega_b~\gg~1$) both
transmission coefficients for the static DB scattering potential
and for the full time-periodic DB scattering potential, decrease
with the breather frequency according to (\ref{Scale}). However,
the effective potential strength $|\beta|$ is larger than
$\beta_0$ and correspondingly the transmission $t_q$ is smaller
compared to the transmission on the static DB scattering
potential. Indeed, we observed this behaviour by direct numerical
simulations (see Fig. \ref{fig9}).

A most peculiar effect is that the presence of ac term in a scattering potential
allows to tune the frequency of a local mode and to obtain the 
resonant suppression of transmission.
Indeed by taking into account the type of diagrams shown in Fig. 3c we obtain
for the renormalized local mode frequency $\tilde{\omega}_L$
\begin{equation}\label{omegaL}
\tilde{\omega}_L^2~=~\omega_{L}^2+\frac{\beta_2^2}{(4\Omega_b-\omega_q)^2-\omega_{L}^2}~~.
\end{equation}
This formula is valid in the limit $\beta_2~\leq~2\sqrt{2}\Omega_b$
\cite{comment1}. 
For the particular case of the DB with frequency $\Omega_b~=~1.5$ and $\beta_2~=~v_{02}/2~=~1.5$ we find 
that
the resonant value of $q~\simeq~2$ is remarkably close to the resonant suppression of $t_q$ around $q~=~2.1$.
Thus a strong dependence of $\omega_L$ on the amplitude of the ac part of DB scattering potential, and 
therefore, on the breather frequency, allows easily to change a position of the resonance. 

Note here, that taking into
account the static part and the second harmonic of the DB
potential allows to obtain a good agreement with the direct
numerical simulations of a scattering by the "full" DB (compare
solid and dashed lines in Fig. \ref{fig10}). It is also interesting to
mention that the static part and the first
harmonic of the DB potential ($k~=~1$) lead to an {\it increase}
of the transmission coefficient compared to the scattering by the
time-averaged DB potential. It is just due to the interplay
between the strengths of different channels of the scattering
potential (see Eq. (\ref{betaRen})).

Thus, we conclude that the presence of a closed channel allowing
absorption and emission of phonons around the center of the
breather leads to strong interference effects which are of
destructive nature in the given example of optical breather.

%

\subsection{The case
$|\Omega_b| < \min{| \omega_q |}$}

\begin{figure}[htb]
\vspace{20pt}
\psfrag{x}{$\hat{x}_n(0)$}
\centerline{\psfig{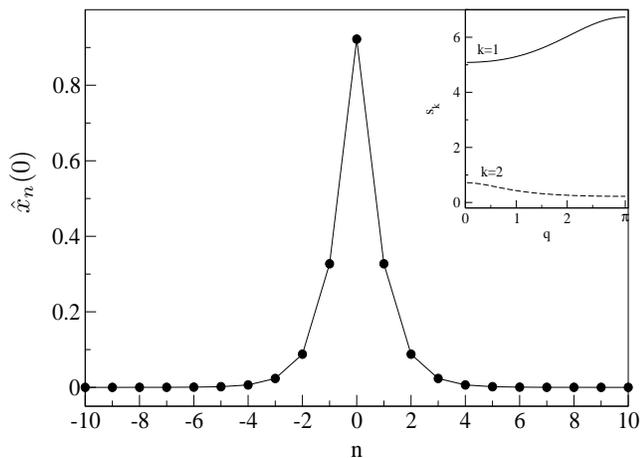}}
\vspace{2pt}
\caption{ The initial displacements of an optical breather with $\Omega_b=0.85$
and $c=0.15$
(velocities are zero).
\\
Inset: $s_{k=1,2}$ versus $q$.
}
\label{x3_profile}
\end{figure}

\begin{figure}[htb]
\vspace{20pt}
\psfrag{v}{$|v_{n,k}|$}
\centerline{\psfig{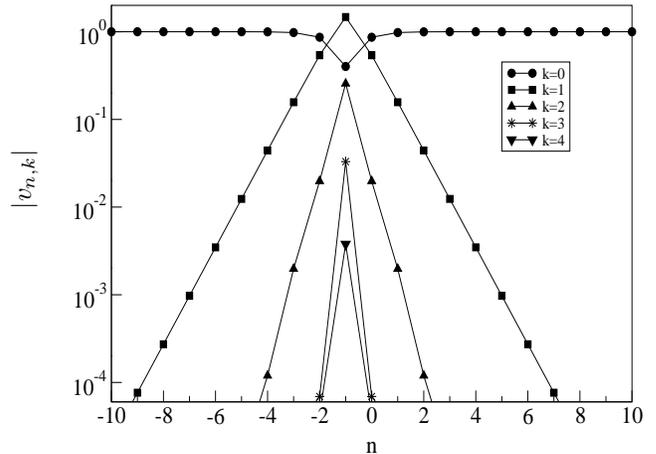}}
\vspace{2pt}
\caption{ Fourier components $v_{nk}$ for different $k$ versus $n$ for the
breather in Fig.\ref{x3_profile}.}
\label{x3_harm}
\end{figure}

\begin{figure}[htb]
\vspace{20pt}
\psfrag{q}{$q$}
\psfrag{k}{\small $q$}
\centerline{\psfig{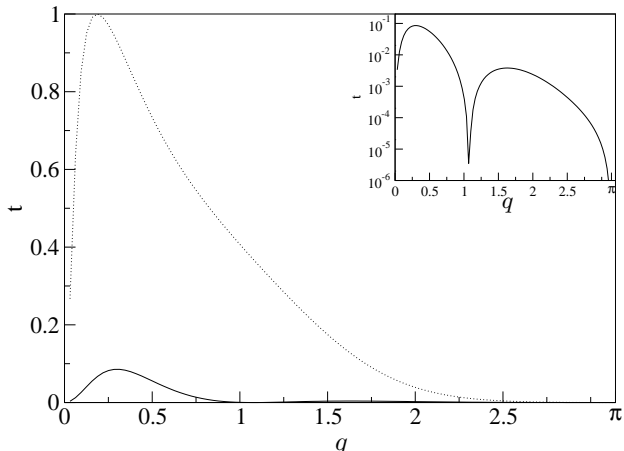}}
\vspace{2pt}
\caption{The numerically calculated dependence of the
transmission coefficient $t_q$ (solid line) on $q$
for the breather frequency $\Omega_b=0.85$ and
$c=0.15$. The dashed line shows the scattering
by the time-averaged DB scattering potential. Inset: The same on a logarithmic scale.}
\label{fig11}
\end{figure}

Here we choose the on-site potential in the form
$V(y)=\frac{1}{2}y^2-\frac{1}{3}y^3$ (note here that the results do not
change if the cubic term has a positive sign) and the interaction term
$W(y)=\frac{c}{2}y^2$.
The breather profile for $\Omega_b=0.85$ and $c=0.15$ is shown
in Fig.\ref{x3_profile}.
The corresponding Fourier components of the scattering potential are plotted
in Fig.\ref{x3_harm}.
In this case the time-averaged DB scattering potential corresponds
to a {\it potential well} in the diagonal terms, i. e. the parameter $\beta_0$
in Eq. (\ref{GrFunctOpt0ch}) has a negative sign. It leads again to the
possibility to obtain a {\it resonance}
as the frequency of an active closed channel matches the frequency of
a localized phonon mode located in the phonon gap. 
Moreover,
at variance to the optical DB with large frequency (preceding subsection),
the number of active closed
channels may now increase substantially. This leads to a
more complicated interference scenario between the open channel and
several closed channels.
Indeed, the estimation of $s_{k=1,2}$ 
shows that the first channel is strongly contributing, and the second closed channel 
can not be neglected either. 
At variance to the previous case for all propagating phonon frequencies 
the $k=1$ closed
channel has a much stronger contribution than the $k=2$ one. However,  we
expect a resonant coupling between the propagating wave and the local mode
through the $k=2$ closed channel, as $| \omega_{q=0} - 2\Omega_b | = 0.7$ is close 
to a local eigenmode of the time-averaged
scattering potential with frequency $\omega_{L} = \pm 0.78$. 
Moreover, the nonresonant processes in 
a strong $k=1$ channel allow to
renormalize $\omega_L$ similarly to the previous case of an optical breather with a large 
frequency.
This resonant effect can be analyzed by making use of
Eqs. (\ref{TransCoeffStat}),
(\ref{GrFunctOpt1})-(\ref{betaRen}). Indeed, due to the
resonance with a localized phonon mode  the Green function
$\tilde G_{-2 \Omega_b+\omega_q}^{(rn)}(0,0)~\gg~1$ for a particular wave number $q_0$.
In this case the parameter $\beta$ goes to infinity as the resonant condition,
$(2\Omega_b-\omega_{q_0})^2-\omega_L^2=0$, is valid
and correspondingly the transmission
coefficient vanishes. \cite{comment2}. As the phonon frequency deviates from the
resonant condition, the parameter $\beta$ decreases and correspondingly the
transmission coefficient reaches a maximum (see Fig. \ref{fig11}). 

In Fig.\ref{fig11} we show that the
time-averaged DB scattering potential provides with perfect transmission at
a certain wave number due to the presence of a quasi-bound
state in the static scattering potential.
At the same time the transmission for the full dynamical problem
shows a maximum value of 0.1, and an additional minimum in
$t(q)$ with actually a full vanishing of transmission. These
patterns are entirely absent in the scattering by the time-averaged potential.
Thus we again conclude that the presence of active closed channels
inside the breather core leads to strong interference effects.
%

\section{Discussion}
In this paper we considered the propagation of small-amplitude phonons through a
nonlinear lattice in the presence of various discrete breathers.
In particular, we obtained the transmission coefficient $t_q$ for the cases of
acoustic and optical discrete breathers, and acoustic rotobreathers.
We have shown that the presence of a DB leads to an effective scattering
potential for phonons and this potential contains both static
(time-averaged) parts and time-dependent periodic parts where the
first and second harmonics are most important.
Moreover, the strength and the width of the scattering potential are
determined by the breather frequency $\Omega_b$.

The static part of the effective DB scattering potential
may fully describe the scattering outcome if the closed channels
weakly contribute (acoustic breather with $\beta=0$). Equally the
leading contribution to a finite transmission may come from a closed
channel (acoustic rotobreather). A very interesting case is realized when
a local mode of the time-averaged scattering potential resonates with a closed
channel (acoustic breather with $\beta \neq 0$, optical breathers).
This resonance leads to a full suppression of transmission at some value of $q$.
The nonresonant closed channels renormalize the local mode frequency and the
corresponding position of the transmission zero in $q$. This effect has
been discussed in several papers in relation to the Fano resonance \cite{swksk01,reichl}.
A detailed explanation of the similarities and differences to the physics
of a Fano resonances is beyond the scope of this work and will be published
separately. It is worthwhile to note here that the location of a zero transmission value
in $q$ is not related to the presence of so-called quasi-bound Floquet states, i.e.
states which by parameter tuning are transformed from a localized state into one
colliding (interacting) with the continous part of the Floquet spectrum. Instead a formally
exact definition of a zero transmission is given in \cite{tcsasf98} through the
asymptotic phase properties of Floquet eigenstates far from the breather center.

The phonon scattering by inhomogeneities can be relevant with respect to
the ongoing discussion
about finite vs. infinite heat conductivity in acoustic chains
with conservation of total mechanical momentum
Ref.~\cite{savgov01,slrlap97,slrlap98,slrlap02}.
The heat conductivity $\kappa$ mediated by noninteracting phonons is determined as
\begin{equation}\label{ThermoCond}
\kappa~\simeq~L \int dq T_q (L)~,
\end{equation}
where $L$ is the size of the system. Here $T_q(L)$ denotes the
transmission of the given system. Thus, obviously in the absence
of inhomogeneities the phonon heat conductivity goes to infinity
($\kappa~\simeq~L$) as the size of the system $L$ increases.
However,  the situation becomes more complex in the presence of
inhomogeneities. In the case, when the inhomogeneities are
randomly distributed along the system, the number of
inhomogeneities $N=nL$, where $n$ is the concentration of defects,
and the total transmission $T_q(L)~\simeq~t_q^N$ with $t_q$ being
the transmission through a given defect. We obtain that the
behaviour of heat conductivity is determined by the values of $q$
where $t_q$ is close to one. Thus, the problem of an infinite heat
conductivity in the limit of a large size $L$ naturally appears in
the systems preserving the total mechanical momentum.

In particular we can apply this general
consideration to the phonon propagation in the presence of
DBs. In the most interesting case of acoustic rotobreathers we obtain
that the transmission coefficient $t_q~\simeq ~1-\alpha q^2$ in the limit of small
wave numbers. The coefficient $\alpha$ is determined by the frequency $\Omega_b$
and the specific choice  of the acoustic rotobreather.
Thus we argue that the mere presence of acoustic rotobreathers does not lead
to a finite heat conductivity and  $\kappa$ is still divergent
in the limit of large $L$ as
\begin{equation}\label{ThermoCond1}
\kappa~\simeq~\sqrt{\frac{L}{n\alpha}}~~.
\end{equation}
Here we assume that the heat is carried by phonons, and did not take into
account phonon-phonon interactions.
Another conclusion drawn from our results for acoustic breathers versus acoustic
rotobreathers is that the $q$-domain around $q=0$ where the transmission is close
to one is by orders of magnitude smaller for acoustic rotobreathers as compared to
acoustic breathers. This region is responsible for the quasiballistic transport
of energy by large wavelength phonons in the hydrodynamic regime and for the appearance
of anomalous heat conductivity. Numerical simulations for systems with acoustic rotobreathers
have to be performed on length and time scales which are thus also orders of magnitude
larger than the corresponding simulations for standard FPU chains.  

Our analysis can be also applied to the electromagnetic wave propagation
through various Josephson transmission lines. Although these systems are
intrinsically dissipative, the dissipation may be rather small, and the
transmission $t_q$ is still determined by the presence of
inhomogeneities like rotobreathers
\cite{etjjmtpo00,pbdaavusfyz00,etjjmabtpo00,aemsfmvfyzjbp01}
or dynamic edge states\cite{FistPage}. Moreover,
in this particular case of Josephson lattices the properties of DBs
can be easily tuned experimentally by changing the external dc bias.
Notice here that local phonon modes located on a
rotobreather play an important role in various switching processes
\cite{aemsfmvfyzjbp01}. A study of an electromagnetic wave propagation may allow
a direct observation of this mode.

Finally, a similar analysis can be carried out for a Schr\"odinger equation in
the presence of artificially applied time-periodic perturbations \cite{comment1}. This case
is important for the description of electron transport through a quantum wire
in the presence of external electromagnetic fields.
\\
\\
\\
{\bf Acknowledgements}
\\
We thank S. Aubry, G. Casati, S. Kim and L. E. Reichl for useful discussions. 
This work was supported by the Deutsche Forschungsgemeinschaft
FL200/8-1.
M. V. F. thank the Alexander von Humboldt Stiftung for partial
supporting this work.
\appendix
\section{Time-averaged scattering potential for acoustic breather}
We take into account the oscillations of two lattice sites in the center of a DB
only. These oscillations evolve exactly in
antiphase with large amplitudes. The effective Hamiltonian is written as
\begin{eqnarray}
E=\frac{\dot{x}_1^2}{2}+\frac{\dot{x}_2^2}{2}+\frac{1}{4}(x_1^4+x_2^4+(x_1-x_2)^4)
\end{eqnarray}
and with $x_1=-x_2$, we arrive at the single oscillator problem
with energy
\begin{eqnarray}\label{enerA}
E=\dot{x}^2+\frac{9}{2}x^4.
\end{eqnarray}
The frequency of oscillation (the breather frequency) is given by
\begin{eqnarray}
\frac{1}{\Omega_b}=\frac{2}{\pi}\int\limits_0^{x_m}\frac{d\,x}{\sqrt{E-\frac{9}{2}x^4}},
\end{eqnarray}
where $x_m=\sqrt[4]{\frac{2E}{9}}$. After integration we obtain
\begin{eqnarray}
\Omega_b=\frac{2\pi\sqrt[4]{\frac{9E}{2}}}{B(\frac{1}{4},\frac{1}{2})},
\end{eqnarray}
where $B(x,y)$ is the $B$-function \cite{isgimr00}. Next we
compute the value of $X=\sqrt{3\langle x^2\rangle}$. We can
express $\langle x^2\rangle$ in terms of the energy
\begin{eqnarray}
\langle
x^2\rangle=\frac{2\Omega_b}{\pi}\int\limits_0^{x_m}\frac{x^2d\,x}{\sqrt{E-\frac{9}{2}x^4}}.
\end{eqnarray}
After some algebra we find $\langle
x^2\rangle=\sqrt{\frac{2E}{9}}\frac{B(\frac{1}{2},
\frac{3}{4})}{B(\frac{1}{4},\frac{1}{2})}$ and
\begin{eqnarray}
X=\sqrt{\frac{2}{3\pi}}\Omega_b.
\end{eqnarray}

\section{Time-averaged scattering potential for acoustic rotobreather}
In order to calculate the time-averaged off-diagonal hopping terms we will take into account
the central site which is in a rotational state, denoted by $\phi$,  and
 two nearest neighbor oscillators (denoted by $\alpha_1$ and $\alpha_2$).
Thus we conserve the total mechanical momentum.
The energy of such a system is
\begin{eqnarray}
E=\dot{\alpha}^2+\frac{\dot{\phi}^2}{2}+2(1-\cos(\phi-\alpha)).
\end{eqnarray}
The two equations of motion are given by
\begin{eqnarray}\label{pend}
  \begin{array}{l}
   \ddot{\phi}=-2\sin(\phi-\alpha),\\
   \ddot{\alpha}=\sin(\phi-\alpha).
  \end{array}
\end{eqnarray}
Introducing new variables $u=\phi+2\alpha$ and $w=\phi-\alpha$ we
rewrite the system (\ref{pend}) as
\begin{eqnarray}
  \begin{array}{l}
   \ddot{u}=0,\\
   \ddot{w}=-3\sin w.
  \end{array}
\end{eqnarray}
The relevant energy part is
\begin{equation}
\tilde{E}=\frac{\dot{w}^2}{2}+3(1-\cos w).
\end{equation}
Now we compute the average coupling between rotational and oscillatory
states for large frequencies
\begin{eqnarray}
\epsilon=\langle\cos w\rangle=\frac{\Omega_b}
{2\pi\sqrt{2\tilde E}}\int\limits_0^{2\pi}\frac{\cos
w\;dw}{\sqrt{1-\frac{3(1-\cos w)}{\tilde E}}}.
\end{eqnarray}
Because $\tilde E \approx \frac{\Omega_b^2}{2}$ we find
\begin{eqnarray}
\epsilon\approx\frac{\Omega_b}{2\pi\sqrt{2\tilde E}}\int\limits_0^{2\pi}
(1+\frac{3(1-\cos w)}{2\tilde E})\cos w\;dw
\end{eqnarray}
which leads to
\begin{eqnarray}
\epsilon=-\frac{3\Omega_b}{4\sqrt{2}\tilde E^{3/2}}
\end{eqnarray}
and finally to
\begin{eqnarray}
\epsilon=-\frac{3}{2\Omega_b^2}.
\end{eqnarray}

\bibliographystyle{unsrt}

\begin{thebibliography}{10}
\bibitem{comment1} Notice here, that a resonant suppression of phonon transmission
can be also realized in an ac driven harmonic lattice in the presence of a static scattering potential. 
In this case the expression (\ref{omegaL}) for the frequency of a local phonon mode is still valid with 
substitutions $2\Omega_b$ to $\omega$, and $\beta_2$ to $A/2$, where $\omega$ and $A$ are correspondingly 
the frequency and amplitude of an external ac drive.
Moreover, a similar resonant suppression of transmission can be observed as an 
electron moving in a periodic potential, 
scatters on a time-dependent potential.
The particular condition for this effect can be written in a simple form: $E \pm 
\omega-\tilde{E}_L~=~0$, where 
$E$ is the electron energy. The renormalized value of the local energy level 
$\tilde{E}_L$ in the limit of small $A$ 
(more precisely $A~\ll~\sqrt{\omega}$ ) is given as
$\tilde{E}_L~=~E_{L} \pm \frac{A^2}{4\omega}$, 
where $E_{L}$ is a local energy level of a static potential.
\bibitem{comment2}
In this particular case the $k=1$ channel strongly contributes.
In order to calculate the renormalized value of 
$\omega_L$ we have to take into account many different diagrams of the kind shown in Fig. 3c.
Although it can be done in the same manner as in the case of a weak renormalization, 
the expression for $\omega_L$ looses its transparency.
\end{thebibliography}

\end{document}